\documentclass
[floatfix,superscriptaddress,secnumarabic,amssymb,amsmath,nobibnotes,aps,prd,showkeys,showpacs,nofootinbib,onecolumn,notitlepage,12pt]{revtex4}%
\usepackage{setspace}
\usepackage{color}
\usepackage{amsmath}
\usepackage{amsfonts}
\usepackage{verbatim}
\usepackage{amssymb}
\usepackage{graphicx,bm}
\usepackage{graphicx}
\usepackage{amsmath}
\usepackage{amssymb}
\usepackage{amssymb}
\usepackage{graphicx,bm}
\usepackage{graphicx}
\usepackage[caption=false]{subfig}%
\providecommand{\U}[1]{\protect\rule{.1in}{.1in}}

\newcommand{\be}{\begin{equation}}
\newcommand{\ee}{\end{equation}}

\newcommand{\mincir}{\raise
-3.truept\hbox{\rlap{\hbox{$\sim$}}\raise4.truept\hbox{$<$}\ }}
\newcommand{\magcir}{\raise
-3.truept\hbox{\rlap{\hbox{$\sim$}}\raise4.truept\hbox{$>$}\ }}

\ifx\pdfoutput\relax\let\pdfoutput=\undefined\fi
\newcount\msipdfoutput
\ifx\pdfoutput\undefined\else
\ifcase\pdfoutput\else
\ifx\paperwidth\undefined\else
\ifdim\paperheight=0pt\relax\else\pdfpageheight\paperheight\fi
\ifdim\paperwidth=0pt\relax\else\pdfpagewidth\paperwidth\fi
\fi\fi\fi
\begin{document}
\title{Lagrangians and Newtonian analogues for Biological Systems}
\author{Andronikos Paliathanasis}
\email{anpaliat@phys.uoa.gr}
\affiliation{Institute of Systems Science, Durban University of Technology, Durban 4000,
South Africa}
\affiliation{Departamento de Matem\'{a}ticas, Universidad Cat\'{o}lica del Norte, Avda.
Angamos 0610, Casilla 1280 Antofagasta, Chile}
\affiliation{National Institute for Theoretical and Computational Sciences (NITheCS), South Africa}
\affiliation{School of Technology, Woxsen University, Hyderabad 502345, Telangana, India}
\author{Kevin Duffy}
\email{kevind@dut.ac.za}
\affiliation{Institute of Systems Science, Durban University of Technology, Durban 4000,
South Africa}
\affiliation{National Institute for Theoretical and Computational Sciences (NITheCS), South Africa}

\begin{abstract}
This study investigates the potential for biological systems to be governed by
a variational principle, suggesting that such systems may evolve to minimize
or optimize specific quantities. To explore this idea, we focus on identifying
Lagrange functions that can effectively model the dynamics of selected
population systems. These functions provide a deeper understanding of
population evolution by framing their behavior in terms of energy-like
variables. We present an algorithm for generating Lagrangian functions
applicable to a family of population dynamics models and demonstrate the
equivalence between two-dimensional population models and a one-dimensional
Newtonian mechanical analog. Furthermore, we explore the existence of
conservation laws for these models, utilizing Noether's theorems to
investigate their implications.

\end{abstract}
\keywords{Lagrangian formilation; population dynamics; cosnervation laws.}\maketitle

\section{Introduction}

Scientists in the early 21st century such as Turchin \cite{Turchin} and
Berryman \cite{Berry} debated the significance of physical laws in population
ecology. They argue that ecology does possess robust, law-like principles that
can guide theoretical and empirical studies, and advocate for formalizing
these principles as laws to strengthen the scientific framework and predictive
power of ecological science. This line of reasoning was pursued further by
Colyvan and Ginsberg \cite{Colyvan} and others who argue that methods and
concepts from physics can help generate new hypotheses and advances in
ecology. Here we consider the concept of physical conservation and initiate a
process of using this concept by considering how population models can be
reformulated to provide conservation laws. Application of these laws could
generate new advances as postulated in \cite{Colyvan}. We must make a
distinction between the use of the word 'conservation' in the natural sciences
as an effort to preserve and protect entities (such as the environment) and
our use of this term as the principle by which the total value of a physical
quantity (such as energy, mass, or momentum) remains inherently constant in a
system. This latter definition is implied throughout.

Physics distinguishes itself from other theories within the realm of applied
mathematics through the presence of the variation principle. In analytic
mechanics, this principle is identified as Hamilton's principle, asserting
that the trajectory followed by a dynamical system between two points in
configuration space ensures the stationary condition of the Action Integral.
The corresponding equations of motion are expressed by the Euler-Lagrange
equations derived from the associated function $L$ which defines the Action
Integral $S=\int Ldt$.

The Lagrangian function, denoted as $L$, serves as a crucial element in
describing dynamical systems and offers a unified framework for addressing a
broad spectrum of physical problems. While Newton's laws of motion are
traditionally formulated in Cartesian coordinates, the Lagrangian approach
enables the systematic application of alternative coordinates. Additionally,
this formulation provides a straightforward method for incorporating
constraints through the Lagrange multiplier approach, extending to the
Hamiltonian formulation, and facilitating entry into Quantum Mechanics via the
quantization process.

In 1918, Emmy Noether published pioneering work establishing a connection
between the existence of conservation laws in dynamical systems and
variational symmetries \cite{noe}. Conservation laws play a significant role
in physics, as they are associated with observable quantities. In essence, the
Lagrangian formulation has opened up new avenues for exploring the dynamics of
nonlinear systems.

From the variational principle we know that the introduction of a total
derivative $\dot{f}=\frac{df}{dt}$, in the Lagrangian $L$, leads to the new
Lagrangian $\bar{L}=L+\dot{f}$ which provides the same set of Euler-Lagrange
equations. Hence, the Lagrangian function is not uniquely defined for a
dynamical system. Indeed, the equation of the free particle $\ddot{x}=0$,
follows from the variation of the Lagrangian function $L_{free}=f\left(
\dot{x}\right)  $, where $f$ is an arbitrary nonlinear function, where we can
see that usual Lagrangian $L=T,~T=\frac{1}{2}\dot{x}^{2}$ to be the Kinetic
energy is only one of all the possible Lagrangians. However, what is common
between all these Lagrangians is that the variational principle provides the
same equation of motion. Thus, a dynamical system may admit more than one
Lagrangian functions, but the application of the variational principle should
lead always to the original dynamical system. 

On the other hand, for the oscillator $\ddot{x}+\omega^{2}x=0$, we usually
consider the Lagrangian function%
\begin{equation}
L=T-V~,\label{as1}%
\end{equation}
in which $V\left(  x\right)  =\frac{1}{2}x^{2}$ is the potential function.
However, this Lagrangian is not unique, see \cite{nucc1}. Nevertheless,
Lagrangian functions of the form (\ref{as1}) are common for the description of
various dynamical systems in Newtonian Mechanics.

The existence of a Lagrangian function is essential for many theories, because
it provides a common fundamental stating point, that is, the variational
principle, for the description of real world problems. Last but not least, the
Lagrangian formulation allows us to apply new tools to understand the
evolution of the dynamical system. Thus, the derivation of a Lagrangian
function is of important interest in many areas of applied mathematics. The
construction of a Lagrangian function from the dynamical system is known as
the inverse problem, and it has drawn the attention of mathematician in the
last decades. There is no one unique approach for the construction of a
Lagrangian function.

In epistemology, two primary approaches are recognized. Empiricism involves
constructing mathematical models for predictions based on observations and
data collected from the real world. On the other hand, rationalism emphasizes
predicting a system's behavior using mathematical frameworks derived from
logical and theoretical principles. This work investigates the potential
existence of a variational principle that governs biological systems,
suggesting that such systems may operate in a way that minimizes or optimizes
a specific quantity. To explore this, we aim to identify Lagrange functions
that can effectively describe the dynamics of certain population models. These
functions provide insights into the underlying mechanisms of population
evolution by framing them in terms of energy-like quantities, revealing
conservation laws, or identifying equilibrium conditions that guide their behavior.

The Lagrangian formulation of biological systems has been explored previously
in the literature. In \cite{ds1,ds2}, Lagrangian functions linear in momentum
were established for certain population models. Subsequently, as demonstrated
in \cite{nuc0}, this method was revealed to be equivalent to the Jacobi Last
multiplier approach, resulting in Lagrangians that are linear in derivatives.
Recent studies, as outlined in \cite{bio1,bio2}, present an alternative
approach to constructing Lagrangians for dynamical systems in population
dynamics. In this approach, the authors focused on two-dimensional dynamical
systems, elevating the system's order to express the biological dynamics in a
Newtonian context and derive corresponding Lagrangian functions. The derived
Lagrangian functions indicated that the studied biological systems are
non-holonomic, introducing constraints in the Euler-Lagrange equations to
replicate the original system.

A Lagrangian formulation can facilitate a stability analysis by examining the
equilibrium conditions in, for example, a predator-prey system correspond to
minima, maxima, or saddle points in the potential energy. Also, it can be used
to calculate action, which is the integral of the Lagrangian over time.
Minimizing the action can provide optimal paths for the system's evolution,
which can be relevant in understanding ecosystem dynamics and managing those
systems. The Lagrangian framework can also be adapted to include more complex
interactions, such as a carrying capacity of the prey or a more complicated
functional response of the predators to the prey population. This can be
achieved by modifying the "kinetic" and "potential" terms in the Lagrangian. A
Lagrangian approach could provide a more extensive bifurcation analysis,
mapping the full parameter ranges where different attractors (equilibria,
limit cycles) exist and their stability for the given biological system.

In our study, we adopt a distinct strategy to address the same problem,
employing a simpler method for deriving Lagrangian functions. Our aim is to
demonstrate that biological systems conform to holonomic dynamical systems.
Notably, this approach allows us to apply Noether's theorems and establish
conservation laws. The paper's structure is outlined as follows.

In Section \ref{sec2} we present the basic mathematical elements which are
applied in this study. Section \ref{sec3} includes the main results of this
work where we employ an algorithm in order to write two-dimensional population
dynamical systems into the form of a Newtonian mechanical analogue. Then we
are able to derive a Lagrangian function and write Noetherian conservation
laws. Finally, in Section \ref{sec4} we discuss our results.

\section{Li\'{e}nard type equation}

\label{sec2}

Consider the so-called Linear-type differential second-order differential
equation%
\begin{equation}
\ddot{x}+f\left(  x\right)  \dot{x}^{2}+\beta\left(  x\right)  \dot{x}%
+\gamma\left(  x\right)  =0, \label{as.02}%
\end{equation}
where $t$ is the independent variable, $x=x\left(  t\right)  $ is the
dependent variable and $\dot{x}=\frac{dx}{dt}$, ~$\ddot{x}=\frac{d^{2}%
x}{dt^{2}}$. Functions $f\left(  x\right)  $, $g\left(  x\right)  $ and
$\gamma\left(  x\right)  $ are arbitrary functions in terms of the dependent variable.

We employ the change of variable $dy=e^{\int f\left(  x\right)  dx}dx$, and in
the new coordinates we eliminate the quadratic term$~f\left(  x\right)
\dot{x}^{2}$. From the chain rule it follows%
\begin{equation}
\dot{y}=e^{\int f\left(  x\right)  dx}\dot{x}~,~\ddot{y}=e^{\int f\left(
x\right)  dx}f\left(  x\right)  \dot{x}^{2}+e^{\int f\left(  x\right)
dx}\ddot{x}.
\end{equation}
Indeed, in the new variables $\left\{  t,y\right\}  $ equation (\ref{as.02})
becomes
\begin{equation}
\ddot{y}+\beta\left(  y\right)  \dot{y}+\hat{\gamma}\left(  y\right)
=0~,~\hat{\gamma}\left(  y\left(  x\right)  \right)  =e^{\int f\left(
x\right)  dx}\gamma\left(  x\right)  . \label{as.03}%
\end{equation}
As a result the quadratic term in Li\'{e}nard equation (\ref{as.02}) can be
eliminated due to effects of the coordinate system.

A natural question that arises is whether there are other transformations that
can eliminate additional terms in equation (\ref{as.03}).

For specific functional forms of $\beta\left(  y\right)  $ and $\hat{\gamma
}\left(  y\right)  $, equation (\ref{as.03}) has the property that it is
maximally symmetric and can be linearized. Indeed, for $\beta\left(  y\right)
=\beta_{0}$ and $\hat{\gamma}\left(  y\right)  =\gamma_{0}y$, equation
(\ref{as.03}) takes the form of an oscillator with a damping parameter
$\beta_{0}$. Now it is known that there exists a point transformation $\left(
t,y\right)  \rightarrow\left(  \tau,z\right)  $, and the equation now reads
$\frac{d^{2}z}{d\tau^{2}}=0$. A similar transformation exists for the
Painlev\'{e}-Ince equation with $\beta\left(  y\right)  =3y$ and $\hat{\gamma
}\left(  y\right)  =y^{3}$. The linearization process is achieved using the
Lie point symmetry approach \cite{olver,ser1}. Nevertheless, it's worth noting
that this is not the only approach, as other transformations, such as nonlocal
transformations, can also be applied.

We use these transformation methods to eliminate $\beta\left(  y\right)  $
from equation (\ref{as.03}) and consider the nonlinear transformation
\begin{equation}
ds=e^{-\int\beta\left(  y\left(  t\right)  \right)  dt}dt.
\end{equation}
In terms of the new independent variable $s$, equation (\ref{as.03}) now reads%
\begin{equation}
y^{\prime\prime}+F\left(  s,y\right)  =0~. \label{as.04}%
\end{equation}
where $y^{\prime}=\frac{dy}{ds}$ and the new function $F\left(  s,y\right)  $
is defined as $F\left(  s,y\right)  =\chi\left(  s\right)  \hat{\gamma}\left(
y\right)  $,~with $\chi\left(  s\right)  =e^{2\int\beta\left(  y\left(
t\right)  \right)  dt}$.

Moreover, if $F\left(  s,y\right)  $ is a linear function on $y$, equation
(\ref{as.04}) can be written in the form of a free particle, by using point
transformations. We proceed with the derivation of Lagrangian functions for
the second-order Li\'{e}nard equation.

\subsection{Lagrangian formulation}

We proceed with the derivation of Lagrangian functions for the second-order
Li\'{e}nard equation.

If functions $\beta\left(  y\right)  $ and $\hat{\gamma}\left(  y\right)  $
are related with the expression
\begin{equation}
\frac{d}{dy}\left(  \frac{\hat{\gamma}\left(  y\right)  }{\beta\left(
y\right)  }\right)  =\alpha\left(  1-\alpha\right)  \beta\left(  y\right)
\,,~\alpha\left(  1-\alpha\right)  \neq0, \label{as.05}%
\end{equation}
then equation (\ref{as.03}) admits the Lagrangian function \cite{nuc2}%
\begin{equation}
L\left(  t,y,\dot{y}\right)  =\left(  \dot{y}+\frac{1}{\alpha}\frac
{\hat{\gamma}\left(  y\right)  }{\beta\left(  y\right)  }\right)  ^{1-\frac
{1}{\alpha}}.
\end{equation}

The latter Lagrangian is autonomous and admits for arbitrary functions, the
Noether symmetry $\partial_{t}$ provides the conservation law%
\begin{equation}
\Phi_{0}=\left(  \dot{y}+\frac{1}{\alpha}\frac{\hat{\gamma}\left(  y\right)
}{\beta\left(  y\right)  }\right)  ^{1-\frac{1}{\alpha}}\frac{\left(
\alpha-1\right)  \beta\left(  y\right)  \dot{y}-\hat{\gamma}\left(  y\right)
}{\alpha^{2}\beta\left(  y\right)  ^{2}}\text{.}%
\end{equation}

However, in the case where functions $\beta\left(  y\right)  $ and
$\hat{\gamma}\left(  y\right)  $ are not related with the expression
(\ref{as.05}) we can work with equation (\ref{as.04}). Conservation law
$\Phi_{0}$ is the \textquotedblleft energy\textquotedblright\ for this
dynamical system. We remark that with the term autonomous we refer to the
property the Lagrangian function do not have an direct dependence to the time
variable, which means that the resulting dynamical system is \textquotedblleft
closed\textquotedblright\ and there exists the conservation law of
\textquotedblleft energy\textquotedblright.

\subsection{Newtonian analogue}

Equation (\ref{as.04}) is a time-independent second-order differential
equation which can describe a Newtonian system with central force. Hence, a
known Lagrangian for this equation of the form (\ref{as1}), given in
\cite{mechanics}, is%

\begin{equation}
L\left(  s,y,y^{\prime}\right)  =\frac{1}{2}y^{\prime2}-V\left(  s,y\right)
~,~F\left(  s,y\right)  =-V_{,y}\text{.}%
\end{equation}
in which $y^{\prime}=\frac{dy}{ds}$.

When $V\left(  s,y\right)  =V\left(  y\right)  $ the latter Lagrangian is
time-independent and admits as a variational symmetry the vector field
$\partial_{s}$. Other symmetries exist for specific functional forms of
$V\left(  y\right)  $; however, the vector field $\partial_{s}$ is the common
symmetry for arbitrary potential function $V\left(  y\right)  $. Hence, from
Noether's second theorem it follows that the quantity%
\begin{equation}
H\equiv\frac{1}{2}y^{\prime2}+V\left(  y\right)  ,
\end{equation}
is a conservation law for equation (\ref{as.04}). Indeed, function $\hat{h}$
is nothing else that the total \textquotedblleft energy\textquotedblright\ for
the closed (autonomous) Newtonian system.

On the other hand conservation laws for potential functions $V\left(
s,y\right)  $ with $V_{,s}\neq0$, are investigated in \cite{karp}.

\section{Biological Systems}

\label{sec3}

In this section, we focus on dynamical systems that describe populations in
biological systems. Specifically, we examine two dynamical systems known as
(i) the Lotka-Volterra system, (ii) the SIR model, and two generalizations,
namely (iii) the host-parasite system and (iv) the Easter Island population
model. Our aim is to demonstrate that the dynamical equations for these models
can be reformulated in an equivalent form of a second-order differential
equation of Li\'{e}nard-type (\ref{as.02}). Subsequently, by applying the
earlier analysis, we can express the Newtonian analogue for each biological
system and derive the corresponding Lagrangian function. In what follows we
provide known models that are the basis of each system considered. As we are
interested in the mathematical derivations of these systems we do not go into
the definitions of variables and parameters which are well known.

\subsection{Lotka-Volterra}

The Lotka-Volterra system is a pair of first-order, non-linear, differential
equations frequently used to model the interactions between two species in a
biological community, where one species is the predator and the other is the
prey. The system is expressed as \cite{bbs1,bbs2}
\begin{align}
\dot{w}_{1} &  =w_{1}\left(  1+bw_{2}\right)  ,\label{vl.01}\\
\dot{w}_{2} &  =w_{2}\left(  A+Bw_{1}\right)  ,\label{vl.02}%
\end{align}
in which $w_{1}$ is the population density for the pray and $w_{2}$ describes
the population density for the predator. Parameter $A$ describes the
predator's death rate.

The latter system is equivalent to the second-order ordinary differential
equation
\begin{equation}
\ddot{w}_{1}-\frac{1}{w_{1}}\left(  \dot{w}_{1}\right)  ^{2}-\left(
A+Bw_{1}\right)  \left(  \dot{w}_{1}-w_{1}\right)  =0. \label{vl.03}%
\end{equation}

Equation (\ref{vl.03}) is of the form of (\ref{as.02}). We introduce the new
variable $w_{1}=e^{W}$, and we eliminate the quadratic term. Thus, equation
(\ref{vl.03}) becomes%
\begin{equation}
\ddot{W}+\left(  A+Be^{W}\right)  \left(  -\dot{W}+1\right)  =0. \label{vl.04}%
\end{equation}
It is Li\'{e}nard type equation of the form (\ref{as.03}) with
\begin{equation}
\beta\left(  W\right)  =-\left(  A+Be^{W}\right)  \,\ \text{and }\hat{\gamma
}\left(  W\right)  =\left(  A+Be^{W}\right)  .
\end{equation}
It follows that $\frac{d}{dW}\left(  \frac{\hat{\gamma}\left(  W\right)
}{\beta\left(  W\right)  }\right)  =0$, which means that in order to construct
a Lagrangian function we should follow the second approach.

Hence, we employ the new independent variable%
\begin{equation}
ds=e^{\int\left(  A+Be^{W}\right)  dt}dt\text{, }%
\end{equation}
such that, equation (\ref{vl.04}) reads%
\begin{equation}
W^{\prime\prime}+\chi\left(  s\right)  \left(  A+Be^{W}\right)  =0~,~\chi
\left(  s\right)  =e^{-2\int\left(  A+Be^{W}\right)  dt},
\end{equation}
where the corresponding Lagrangian is
\begin{equation}
L\left(  s,W,W^{\prime}\right)  =\frac{1}{2}W^{\prime2}-\chi\left(  s\right)
\left(  AW+Be^{W}\right)  .
\end{equation}

We observe that parameter $b$ does not appear in the Lagrangian function
neither in the reduced system. That is not a surprise because value $b$ is not
essential and it can be eliminated. Indeed, by assuming $w_{2}=b\bar{w}_{2}$
parameter $b$ is omitted from the system (\ref{vl.01}), (\ref{vl.02}).

\subsection{SIR model}

The Susceptible, Infectious, or Recovered (SIR) model describes the dynamics
of an infectious disease within a population. The simple SIR model is the
following one \cite{bbs3}
\begin{align}
\dot{w}_{1}  &  =-w_{1}w_{2},\\
\dot{w}_{2}  &  =w_{1}w_{2}-aw_{2},
\end{align}
in which $w_{1}$ is the stock of susceptible population and $w_{2}$ is the
stock of infected population.

The dynamical system is equivalent with the second-order differential equation%
\begin{equation}
\ddot{w}_{1}-\frac{1}{w_{1}}\left(  \dot{w}_{1}\right)  ^{2}+\left(
a-w_{1}\right)  \dot{w}_{1}=0,
\end{equation}
or%
\begin{equation}
\ddot{W}+\left(  \alpha-e^{W}\right)  \dot{W}=0. \label{sir.04}%
\end{equation}

As before, condition (\ref{as.05}) is not valid. Thus, we apply the change of
variable $ds=e^{-\int\left(  \alpha-e^{W}\right)  dt}dt$, where equation
(\ref{sir.04}) takes the form of the free particle, that is,%
\begin{equation}
W^{\prime\prime}=0.
\end{equation}

One of the many Lagrangians is the
\begin{equation}
L\left(  s,W,W^{\prime}\right)  =\frac{1}{2}W^{\prime2},
\end{equation}
and the conservation law, momentum $W^{\prime}=const$, and the energy $\hat
{h}=\frac{1}{2}W^{\prime2}.$

Let us now reconstruct the conservation law for the original dynamics. Indeed,
$W^{\prime}=const$, results in%
\begin{equation}
\dot{W}e^{-\int\left(  \alpha-e^{W}\right)  dt}=const
\end{equation}
or%
\begin{equation}
\frac{\dot{w}_{1}}{w_{1}}e^{-\int\left(  \alpha-w_{1}\right)  dt}=const.
\end{equation}

\subsection{SIR\ model II}

For cancer a common model describes the interaction between infected and
uninfected tumor cells, the amount of virus and the change in the size of the
tumor over time. The system is expressed by the following\ modified SIR model
\cite{warz0}%
\begin{align}
\dot{w}_{1}  &  =\lambda-\delta w_{1}-w_{1}w_{2},\\
\dot{w}_{2}  &  =w_{1}w_{2}-aw_{2}.
\end{align}

In a similar procedure as before we can write the second-order differential
equation
\[
\ddot{W}-\left(  e^{W}-\lambda e^{-W}-a\right)  \dot{W}-\left(  1-ae^{-W}%
\right)  \left(  \delta-\lambda e^{-W}\right)  =0,
\]
where $w_{1}=e^{W}$.

Therefore, the Newtonian analogue is
\begin{equation}
W^{\prime\prime}-\chi\left(  s\right)  \left(  \left(  1-ae^{-W}\right)
\left(  \delta-\lambda e^{-W}\right)  \right)  =0~,~\chi\left(  s\right)
=e^{2\int\left(  e^{W}-\lambda e^{-W}-a\right)  dt},
\end{equation}
and the corresponding Lagrangian function%
\[
L\left(  s,W,W^{\prime}\right)  =\frac{1}{2}W^{\prime2}+\chi\left(  s\right)
\left(  \left(  a\delta+\lambda\right)  e^{-W}-\frac{1}{2}a\lambda
e^{-2W}+\delta\right)  =0.
\]

\subsection{SIR\ model III}

A modified SIR dynamical model which describes a growing tumor and an
oncolytic virus population over time is introduced in \cite{warz}. This
two-dimensional dynamical system is given by%
\begin{align}
\dot{w}_{1}  &  =rw_{1}\left(  1-\frac{w_{1}+w_{2}}{\omega}\right)  -\delta
w_{1}-w_{1}w_{2},\\
\dot{w}_{2}  &  =w_{1}w_{2}-aw_{2}.
\end{align}

Thus, the equivalent second-order equation becomes
\begin{equation}
\ddot{W}+\frac{\left(  e^{W}\left(  r-\omega\right)  +a\omega\right)  }%
{\omega}\dot{W}-\frac{\left(  \left(  e^{W}-a\right)  \left(  e^{W}%
r-\omega\left(  r-\delta\right)  \right)  \right)  }{\omega}=0,
\end{equation}
in which $w_{1}=e^{W}$.

After the nonlocal transformation $t\rightarrow s$, the latter equation reads%
\begin{equation}
W^{\prime\prime}-\chi\left(  s\right)  \frac{\left(  \left(  e^{W}-a\right)
\left(  e^{W}r-\omega\left(  r-\delta\right)  \right)  \right)  }{\omega
}=0~,~\chi\left(  s\right)  =e^{-2\int\frac{\left(  e^{W}\left(
r-\omega\right)  +a\omega\right)  }{\omega}dt},
\end{equation}
where now the corresponding Lagrangian is%
\begin{equation}
L\left(  s,W,W^{\prime}\right)  =\frac{1}{2}W^{\prime2}+\frac{\chi\left(
s\right)  }{\omega}\left(  e^{W}\left(  r\left(  e^{W}-2a-2\omega\right)
+2\delta\omega\right)  +a\left(  \delta-r\right)  \omega W\right)  \text{.}%
\end{equation}

\subsection{Host-parasite model}

The host-parasite model is a two-dimensional system expressed by the following
set of equations \cite{bbs4}%
\begin{align}
\dot{w}_{1}  &  =w_{1}\left(  1-\frac{w_{2}}{K}\right)  ,\\
\dot{w}_{2}  &  =aw_{2}\left(  1-B\frac{w_{2}}{w_{1}}\right)  ,
\end{align}
where $w_{1}$ is the density for the host population and $w_{2}$ is the
density for the parasites.

Now the second-order ordinary differential equation is%
\begin{equation}
\ddot{w}_{2}-\frac{2}{w_{2}}\left(  \dot{w}_{2}\right)  ^{2}+\left(
2-\frac{w_{2}}{K}\right)  \dot{w}_{2}+\left(  \frac{w_{2}}{K}-1\right)
w_{2}=0,
\end{equation}
or%
\begin{equation}
\ddot{W}+\left(  2-\frac{1}{KW}\right)  \dot{W}+\left(  W-\frac{1}{K}\right)
=0,~~W=\frac{1}{w_{2}}.
\end{equation}

It is easy to see that condition (\ref{as.05}) is not satisfied. Hence, we
introduce the new independent variable%
\begin{equation}
ds=e^{-\int\left(  2-\frac{1}{KW}\right)  dt}dt
\end{equation}
resulting in the equations%
\begin{equation}
W^{\prime\prime}+\chi\left(  s\right)  \left(  W-\frac{1}{K}\right)
=0~,~\chi\left(  s\right)  =e^{2\int\left(  2-\frac{1}{KW}\right)  dt},
\label{sd11}%
\end{equation}
and the corresponding Lagrangian function is
\begin{equation}
L\left(  s,W,W^{\prime}\right)  =\frac{1}{2}W^{\prime2}-\chi\left(  s\right)
\left(  \frac{W^{2}}{2}-\frac{1}{K}W\right)  .
\end{equation}

Equation (\ref{sd11}) is nothing else than that of the forced oscillator
\cite{leach}. Indeed, equation (\ref{sd11}) is a linear second-order
differential equation and it has the property to be maximally symmetric
\cite{leach}, as a result the equation can be written in the equivalent form
of the free particle under a change of variables.

The solution is expressed as $W=W_{h}+W_{s}$, where $W_{s}$\ is a special
solution of (\ref{sd11}) and $W_{h}$\ solves the homogeneous equation%

\begin{equation}
W_{h}^{\prime\prime}+\chi\left(  s\right)  W_{h}=0. \label{sd.11a}%
\end{equation}
Then under the change of variables
\begin{equation}
Q=\frac{W_{h}}{\rho}~,~T=\int\rho^{-2}\left(  s\right)  ds
\end{equation}
the homogeneous equation reads%
\begin{equation}
\frac{d^{2}Q}{dT^{2}}+Q=0 \label{sd.12}%
\end{equation}
in which $\rho$\ solves the equation%
\begin{equation}
\rho^{\prime\prime}+\chi^{2}\left(  s\right)  \rho=\frac{1}{\rho^{3}}=0.
\end{equation}

Hence, a conservation law for equation (\ref{sd.12}) is the Hamiltonian
function $h=\left(  \frac{dQ}{dT}\right)  ^{2}-Q^{2}$. However, it is
important to mention that equation (\ref{sd.11a}) admits the conservation law
known as the Lewis-Invariant \cite{leach}%
\begin{equation}
I=\frac{1}{2}\left(  \left(  \rho W^{\prime}-\rho^{\prime}W\right)  ^{2}%
+\frac{W^{2}}{\rho^{2}}\right)  .
\end{equation}
This conservation law is essential to describe the integrability properties of
the above system.

\subsection{Easter Island population model}

The Easter Island population model describes the dynamics of population growth
expressed by the logistic equations \cite{bb1}%
\begin{align}
\dot{w}_{1}  &  =cw_{1}\left(  1-\frac{w_{1}}{K}\right)  -hw_{2},\\
\dot{w}_{2}  &  =aw_{2}\left(  1-\frac{w_{2}}{w_{1}}\right)  .
\end{align}
Variable $w_{1}$ is the amount of resources and $w_{2}$ is the population.
Constant $K$ is the carrying capacity, $a$ the growth rate of the population,
$c$ is the growth rate of the resources and $h$ the harvesting constant.

The Lagrangian formulation of the latter model by using the multiplier
approach was investigated before in \cite{nuc3}.

The equivalent second-order differential equation is
\begin{equation}
\ddot{W}+\left(  a+c+2h\right)  \dot{W}+\frac{a+h}{K}\left(  \left(
c+h\right)  W-cW^{\frac{h}{a+h}}\right)  =0,
\end{equation}
or equivalent%
\begin{equation}
W^{\prime\prime}+\frac{1}{s^{2}}\frac{a+h}{K}\left(  \left(  c+h\right)
W-cW^{\frac{h}{a+h}}\right)  =0,
\end{equation}
where $w_{2}=W^{-\frac{a}{a+h}}$.

Finally, the Lagrangian function is
\begin{equation}
L\left(  s,W,W^{\prime}\right)  =\frac{1}{2}W^{\prime2}-\frac{1}{s^{2}}%
\frac{a+h}{K}\left(  \frac{\left(  c+h\right)  }{2}W^{2}+\frac{\left(
a+h\right)  c}{h}W^{1+\frac{h}{a+h}}\right)  =0.
\end{equation}

In the special limit where $a+h=0$, the latter dynamical system reduces to
that of the free particle
\[
W^{\prime\prime}=0\text{, }%
\]
similar to that of the SIR model studied before.

At the other limit, where $c+h=0$, \ the Lagrangian function reads%
\begin{equation}
L\left(  s,W,W^{\prime}\right)  =\frac{1}{2}W^{\prime2}-\frac{1}{s^{2}}%
\frac{a+h}{K}\left(  \frac{\left(  a+h\right)  c}{h}W^{1+\frac{h}{a+h}%
}\right)  =0.
\end{equation}
which is a central forces problem with varying mass, where the invariant
functions have been investigated before in \cite{mit1}.

\section{Conclusions}

\label{sec4}

We addressed six biological systems that depict the dynamics of
two-dimensional population models. Our approach involves presenting an
algorithm tailored to articulate the Newtonian mechanical analogues for this
specific class of models. Furthermore, we establish that these models can be
derived from a variational principle by formulating their corresponding
Lagrangian functions.

The distinctive property exhibited by this family of models facilitates the
development of conservation laws through the application of Noether's
theorems. The identification of the Lagrangian function introduces novel
avenues for exploring population dynamics. Moreover, we highlight the
potential for quantizing the population dynamical systems.

As an example of how this might be applied further, consider a predator-prey
system using differential equations as describe here. By applying the
Lagrangian approach these interactions are reformulate in terms conservation
principles. In practice, one might define the "kinetic" part to represent the
growth potential of the prey population and the "potential" part as the
limitations imposed by predators and environmental factors. By analysing this
system through the Lagrangian framework, conservation laws can provide
invariant quantities in the system, such as the total energy (representing a
balance of population dynamics). This could help ecologists predict population
changes more accurately and also manage ecosystems more effectively by
identifying critical points that need intervention to maintain ecological
balance. For example, controlling the number of predators to prevent the prey
population from crashing or exploding, which could have broader implications
for ecosystem stability.

At this point it is worth to mention that another physical theory that is used
for the study of biological models is the thermodynamic approach
\cite{th1,th2}. The concept of energy and entropy are defined such that two
write the thermodynamical laws in population dynamical models. This approach
differs from what we followed in this study. In particular in this work we
assumed the dynamical variables to be described as point-like particles, while
in thermodynamical approach the biological system is considered as multi-body
system, in a mathematical approach similar to that of physical systems. Due to
this consideration, the variation principle we considered in this study is
that of the minimum length; nevertheless in thermodynamic approach there exist
another generalized variation principle known as Onsager's variational
principle \cite{on1}, which has been examined in population dynamics
\cite{on2,on3}. Onsager's variational principle what is minimized is the
entropy production. In contrary to the Hamiltonian principle which describes
equilibrium systems, Onsager's variational principle has been proposed to
describe nonequlibrium dynamical systems.

In a forthcoming investigation, we aim to investigate further the variation
principle of population dynamics models, and to find common futures.
Additionally, we aspire to extend the algorithm to construct Lagrangian
functions for higher-dimensional biological models.

\bigskip

\textbf{\bigskip Data Availability Statement:} No data made in this data.

\textbf{Acknowledgements}

AP thanks the support of VRIDT through Resoluci\'{o}n VRIDT No. 096/2022,
Resoluci\'{o}n VRIDT No. 098/2022. KD was funded by the National Research
Foundation of South Africa, Grant number 131604. The authors thanks Prof. Cang
Hui for a fruitfull discussion and the Stellenbosch University while part of
this work was carried out. AP thanks Dr. Amlan Halder and the Woxsen
University for the hospitality provided while part of this work was carried out.

\end{document}